\documentclass[fleqn,usenatbib,useAMS]{mnras}

\usepackage{graphicx}	
\usepackage{amsmath}	
\usepackage{amssymb}	
\usepackage{multicol}        
\usepackage{bm}		
\usepackage{pdflscape}	

\usepackage{adjustbox}
\usepackage{pifont}
\usepackage{float}
\usepackage{graphicx}	
\usepackage{multirow}
\usepackage{caption}
\usepackage{subcaption}
\usepackage{booktabs}
\usepackage{times}
\usepackage{xcolor}
\usepackage{ulem}
\usepackage{hyperref}

\usepackage[T1]{fontenc}
\usepackage{ae,aecompl}

\usepackage{newtxtext,newtxmath}

\title[PLATYPUS: A Survey of Debris Discs at 8.8mm]{Four new PLanetesimals Around TYpical and Pre-main seqUence Stars (PLATYPUS) Debris Discs at 8.8mm}

\author[Brodie J. Norfolk]{
Brodie J. Norfolk$^{1}$\thanks{Contact e-mail: \href{mailto:bnorfolk@swin.edu.au}{bnorfolk@swin.edu.au}}, Sarah T. Maddison$^{1}$, Jonathan P. Marshall$^{2,3}$, Grant M. Kennedy$^{4,5}$, \newauthor
Gaspard Duch\^{e}ne$^{6,7}$, David J. Wilner$^{8}$, Christophe Pinte$^{9,10}$, Attila Mo\'or$^{11,12}$, \newauthor
Brenda Matthews$^{13,14}$, P\'eter \'Abrah\'am$^{11,12}$, \'Agnes K\'osp\'al$^{11,12,15}$, Nienke van der Marel$^{13,14}$
\\
$^{1}$Centre for Astrophysics and Supercomputing (CAS), Swinburne University of Technology, Hawthorn, Victoria 3122, Australia\\
$^{2}$Academia Sinica, Institute of Astronomy and Astrophysics, 11F Astronomy-Mathematics Building, NTU/AS campus,\\ No. 1, Section 4, Roosevelt Rd., Taipei 10617, Taiwan\\
$^{3}$Centre for Astrophysics, University of Southern Queensland, Toowoomba, QLD 4350, Australia\\
$^{4}$Department of Physics, University of Warwick, Gibbet Hill Road, Coventry, CV4 7AL, UK\\
$^{5}$Centre for Exoplanets and Habitability, University of Warwick, Gibbet Hill Road, Coventry CV4 7AL, UK\\
$^{6}$Astronomy Department, University of California, Berkeley, CA 94720, USA\\
$^{7}$Universit\'e Grenoble Alpes / CNRS, Institut de Plan\'etologie et d'Astrophysique de Grenoble, 38000 Grenoble, France\\
$^{8}$Harvard-Smithsonian Center for Astrophysics, 60 Garden Street, Cambridge, MA 02138, USA\\
$^{9}$Monash Centre for Astrophysics (MoCA) and School of Physics and Astronomy, Monash University, Clayton Vic 3800, Australia\\
$^{10}$Univ. Grenoble Alpes, CNRS, IPAG, F-38000 Grenoble, France\\
$^{11}$Konkoly Observatory, Research Centre for Astronomy and Earth Sciences, E\"otv\"os Lor\'and Research Network (ELKH),\\ 
H-1121 Budapest, Konkoly-Thege Mikl\'os \'ut 15--17, Hungary\\
$^{12}$ELTE E\"otv\"os Lor\'and University, Institute of Physics, P\'azm\'any P\'eter s\'et\'any 1/A, 1117 Budapest, Hungary\\
$^{13}$Department of Physics \& Astronomy, University of Victoria, Victoria, BC V8P 5C2\\
$^{14}$Herzberg Astronomy \& Astrophysics Programs, National Research Council of Canada, 5071 West Saanich Road, Victoria BC V9E 2E7, Canada\\
$^{15}$Max Planck Institute for Astronomy, K\"onigstuhl 17, 69117 Heidelberg, Germany}

\date{Last updated: March 2021}

\pubyear{2021}

\begin{document}
\label{firstpage}
\pagerange{\pageref{firstpage}--\pageref{lastpage}}
\maketitle

\begin{abstract}

Millimetre continuum observations of debris discs can provide insights into the physical and dynamical properties of the unseen planetesimals that these discs host.  The material properties and collisional models of planetesimals leave their signature on the grain size distribution, which can be traced through the millimetre spectral index. We present 8.8~mm observations of the debris discs HD~48370, CPD-72~2713, HD~131488, and HD~32297 using the Australian Telescope Compact Array (ATCA) as part of the PLanetesimals Around TYpical Pre-main seqUence Stars (PLATYPUS) survey. We detect all four targets with a characteristic beam size of $5''$ and derive a grain size distribution parameter that is consistent with collisional cascade models and theoretical predictions for parent planetesimal bodies where binding is dominated by self-gravity. We combine our sample with 19 other millimetre-wavelength detected debris discs from the literature and calculate a weighted mean grain size power law index which is close to analytical predictions for a classical steady state collisional cascade model. We suggest the possibility of two distributions of $q$ in our debris disc sample; a broad distribution (where $q \sim 3.2-3.7$) for "typical" debris discs (gas-poor/non-detection), and a narrow distribution (where $q < 3.2$) for bright gas-rich discs. Or alternatively, we suggest that there exists an observational bias between the grain size distribution parameter and absolute flux which may be attributed to the detection rates of faint debris discs at \(\rm \sim\)cm wavelengths.
\end{abstract}




\section{Introduction}
Debris discs are the final stage of protoplanetary disc evolution \citep{2011ARA&A..49...67W,2015Ap&SS.357..103W}. The majority of primordial gas has either accreted onto the star/companions or been blown away by photoevaporative winds, and the remaining dust is replenished through ongoing collisions between dust-producing planetesimals, i.e. asteroids and comets \citep{2008ARA&A..46..339W,2014prpl.conf..521M}.

Collisions in the disc are driven by planetesimal stirring that is triggered by either the interaction with smaller bodies that excite the belt \citep{2002ApJ...577L..35K,2008ApJS..179..451K,2018MNRAS.479.3300K}, or by the dynamical influence of fully formed planets \citep{2009MNRAS.399.1403M}. The size distribution of the grains produced by these collisions provides insight into the different physical and dynamical properties of the invisible parent planetesimals. The original collisional cascade model was formulated by \citet{1969JGR....74.2531D}, who used a power-law grain size distribution \( dn(a) \propto a^{-q} da\), and determined \(q = 3.5\) for grains with constant tensile strength and velocity dispersion. More recently, this standard model has been improved upon to include grain-size dependant tensile strengths \citep{2005Icar..173..342P} and velocity distributions \citep{2012ApJ...749...14G,2012ApJ...747..113P} which result in a range of the grain size distribution exponent $q$ between 3 and 4. This theoretically estimated range of $q$ is supported by a number of millimetre wavelength observations \citep{2012A&A...539L...6R,2015ApJ...813..138R,2016ApJ...823...79M, 2017MNRAS.468.2719M,2018ApJ...855...56W,2020AJ....159..288M}. To date, numerical modelling of observations has constrained the parameter to \( 3.2 < q < 3.8 \) for various grain materials \citep{2018ARA&A..56..541H,2020A&A...641A..75L}. 

The underlying properties of parent planetesimals remains unknown. Observing multiple discs across a range of spectral types and ages can help address this. To further constrain the material properties of dusty debris discs, multi-wavelength observations in the millimetre regime are required to determine the millimetre spectral index, \(\alpha_{\rm mm}\) \citep{1990AJ.....99..924B,2012ApJ...749...14G}. \(\alpha_{\rm mm}\) is a function of the dust emissivity and can be used to constrain the collisional state of the disc \citep{2010RAA....10..383K,2012A&A...539L...6R,2015ApJ...813..138R}. Despite their faint emission, observing at longer wavelengths (\(\rm \sim\)1~cm) provides a better constraint on the mm spectral index (a long lever arm) and is effectively in the Rayleigh-Jeans regime for typical disk temperatures. Here, we present new results from PLATYPUS, an ongoing survey of debris discs at 8.8~mm with the Australian Telescope Compact Array (ATCA) \citep{2015ApJ...813..138R, 2017MNRAS.468.2719M}. The targets in this work are four relatively young debris disc host stars spanning a broad range of stellar luminosities that, including sources from the previously largest disc comparison survey \citep{2020A&A...641A..75L}, increases the number of systems with measured values to 22. 

\section{The Sample}
The PLATYPUS sample are selected to have (1) declinations below 20 degrees in order to be observable with the ATCA array, (2) complementary  ALMA observations  at 1.3~mm, and (3) are relatively compact to maximize surface brightness sensitivity. In this work we add an additional four debris discs not previously observed at long wavelengths that span a range of spectral types and are comparatively young (50~Myr or less).  
We summarise the relevant stellar properties of the four  sources in Table~\ref{tab:platypus_sample}.

\begin{table}
\caption{Stellar properties of our four new PLATYPUS sources.}
\label{tab:platypus_sample}
\centering
\begin{adjustbox}{width=0.48\textwidth}
\begin{tabular}{cccccc} \hline
source & RA & dec & distance (pc) & SpType & age (Myr) \\ \hline
HD 48370 & 06:43:01 & -02:53:19 & 36.07$\pm$0.07 & G8 & 20-50 \\
CPD-72 2713 & 22:42:48 & -71:42:21 & 36.66$\pm$0.03 & K7-M0  & 24\\
HD 131488 & 14:55:08 & -41:07:13 & 155$\pm$2 & A1 & 15 \\
HD 32297 & 05:02:27 & +07:27:39 & 133$\pm$1 & A5-A6 & 15-45 \\
\hline
\end{tabular}
\end{adjustbox}
\end{table}

\subsection{HD~48370}
HD~48370 is a G8~V star \citep{2008hsf2.book..757T} at a distance of \(36.07 \pm 0.07\)~pc \citep{2018A&A...616A...1G} and has an estimated age of \(\sim\)20--50~Myr \citep{2008hsf2.book..757T}. A peak radius for the disc to be at \(\sim\)90~au using spatially resolved \textit{Herschel} images \citep{2016ApJ...826..123M,2021MNRAS.501.6168M}. Using archival 1.3~mm ALMA observations (Project ID: \(\rm 2016.2.00200S\)) of HD~48370, we fit a simple Gaussian to the observatory-calibrated visibilities and derive an integrated flux of \(\rm 5.0 \pm 0.5\)~mJy. 

\subsection{CPD-72 2713}
CPD-72~2713 (CPD-72) is a late-type star with a derived spectral type of K7--M0 \citep{2006A&A...460..695T,2013ApJS..208....9P,2014MNRAS.443.2561G}. It resides at a distance of \(36.66 \pm 0.03\)~pc \citep{2018A&A...616A...1G} and is a member of the \(\rm \approx\) 24~Myr old \(\rm \beta\) Pic moving group \citep{2006A&A...460..695T,2015MNRAS.454..593B,2018MNRAS.475.2955L,2018ApJ...856...23G}. With new 1.33~mm ALMA observations, \citet{2020AJ....159..288M} estimates the outer radius of the cold debris disc surrounding the host star to be 140\(\rm \pm\)14~au.

\subsection{HD~131488}
HD~131488 is an A2 type star \citep{2013ApJ...778...12M} residing at a distance of \(155 \pm 2\)~pc \citep{2018A&A...616A...1G} and is \(\rm \sim\) 15~Myr old \citep{2002AJ....124.1670M,2012ApJ...746..154P}. It is a member of the Upper Centaurus Lupus moving group in the Sco-Cen association \citep{2011MNRAS.416.3108R}. By analysing spatially resolved ALMA continuum observations, \citet{2017ApJ...849..123M} derived a disc radius of $\sim$0\farcs57 ($\sim$88\,au). They also found that the disc harbours a substantial amount of CO gas ($\sim$0.1\,M$_\oplus$, considering the Gaia DR2 based distance of the object).

\subsection{HD~32297}
HD~32297 is an A5~V or A6~V type star \citep{2009ApJ...702..318D} at a distance of \(133 \pm 1\)~pc \citep{2018A&A...616A...1G}. \citet{2005ApJ...635L.169K} estimates an age less than 30~Myr, whereas \citet{2020AJ....160...24E} derived a range of 15--45~Myr. \citet{2018ApJ...869...75M} fit the visibilities of high-resolution 1.3~mm ALMA observations and derive the inner edge of the planetesimal belt to be 76\(\rm \pm\)8\,au and the inner edge of the disc halo to be 122\(\rm \pm\)3\,au, in agreement with previous Keck/NIRC2 imaging \citep{2012ApJ...757...28C}. They also constrain the outer edge of the halo to 440\(\rm \pm\)32~au, closely matching estimates from \textit{HST} images \citep{2005ApJ...629L.117S}. \citet{2020AJ....159..251D} found the disk to be extremely symmetric in scattered light, with disc morphology in reasonable agreement with the ALMA results from \citet{2018ApJ...869...75M}. HD~32297 is considered a CO-rich disc \citep{2016MNRAS.461.3910G} with a mass of \(\rm 7.4\times10^{-2}~M_{\oplus}\). This is the second most massive after HD131488, see \citet{2018ApJ...869...75M} and \citet{2019ApJ...884..108M}, and is one of four CO-rich debris discs around \(\rm \sim\)~30-40~Myr A-type stars found to date, joining HD~21997 \citep{2011ApJ...740L...7M} and 49~Ceti \citep{2012ApJ...758...77Z}. 

\section{ATCA Observations and Data Reduction} 
We used the ATCA radio telescope to conduct our survey at 34~GHz (project code C2694, PI: Maddison). The Compact Array Broadband Backend \citep[CABB,][]{2011MNRAS.416..832W} provides observations with two bands that contain \(\rm 2048 \times 1\)~MHz channels, which we centred at 33~GHz and 35~GHz. Observations were conducted in the hybrid H214 array configuration with antenna 6 flagged due to the increased phase noise on long baselines. This sets an effective baseline range from 92~m to 247~m. The synthesised beam for each observation is detailed in Table~\ref{tab:atca_results}. The weather during the observations varied for each science target, and the seeing monitor RMS path length noise for each observation is summarised in Table \ref{tab:atca_results}.

The science targets were observed with a repeated sequence of 10~min on-source integration and 2~min integration of the gain/phase calibrator. The bandpass and flux calibrators were observed for \(\sim\)15~min and pointing checks were made on the phase calibrator every \(\sim\)60--90~min. All observational and calibration details are summarised in Table~\ref{tab:atca_results}.

\begin{table*}
\caption{PLATYPUS survey results and observing log.}
\label{tab:atca_results}
\centering
\begin{adjustbox}{width=0.99\textwidth}
\begin{tabular}{cccccccccc}
\hline
\multirow{2}{*}{Source} & \multirow{2}{*}{\begin{tabular}[c]{@{}c@{}} Flux \\ Density \\ (\(\rm \mu  Jy\)) \end{tabular}} & \multirow{2}{*}{\begin{tabular}[c]{@{}c@{}} \(1\sigma\) \\ (\(\rm \mu  Jy\))\end{tabular}} & \multirow{2}{*}{\begin{tabular}[c]{@{}c@{}} \(\theta_{\rm beam}\) \\ ($\rm \arcsec \times \arcsec$)\end{tabular}} & \multirow{2}{*}{Date} & \multirow{2}{*}{\begin{tabular}[c]{@{}c@{}}Obs. \\ Time\\ (min)\end{tabular}} & \multicolumn{3}{c}{Calibrators} & \multirow{2}{*}{\begin{tabular}[c]{@{}c@{}}Ave. \\ Path Length\\ rms (\(\rm \mu m \))\end{tabular}} \\ \cline{7-9} & &&&& & Band-pass & Flux & \begin{tabular}[c]{@{}c@{}}Gain/\\ Phase\end{tabular} \\  \hline
HD~48370 & 70.0 & 10.8 & 5.61 x 5.01 & 2019 Mar 01-02 & 225 & 1253-055 & 1934-638 & 0639-032 & 255 \\
CPD-72 2713 & 95.9 & 16.1 & 5.74 x 5.40 & 2019 Mar 02  & 225 & 1921-293 & 1934-638 & 2229-6910 & 329\\
HD~131488 & 59.5 & 12.4  & 5.78 x 4.51 & 2019 Mar 02  & 430 & 1253-055 & 1934-638 & 1451-400 & 111\\
HD~32297 & 56.2 & 16.7 & 6.37 x 4.73 & 2019 Mar 01-02  & 450 & 0003-066 & 1934-638 & 0454+066 & 282\\ \hline
\end{tabular}
\end{adjustbox}
\end{table*}
The data was processed using the software package {\sc Miriad} \citep{1995ASPC...77..433S} and followed the standard procedure which involved: correcting for the frequency-dependent gain using the {\sc Miriad} task \textit{mfcal}; then using the flux density of the ATCA primary flux calibrator, 1937-638, to re-scale the visibilities measured by the correlator using the miriad task \textit{mfcal} with the option \textit{nopassol} set; and correcting for the gain of the system's time variable properties due to changing conditions using the {\sc Miriad} task \textit{gpcal}. To reduce the noise in our data while maintaining as complete an observational track as possible, we flagged all data with a seeing monitor RMS path length noise above 400~$\rm \mu$m using \textit{uvflag}, and calibrator amplitude readings that deviated more than 10\% from the mean flux using \textit{blflag}. Any unusual spikes seen in the channel vs. amplitude or the channel vs. phase plots were also flagged using \textit{uvflag}. We adopt a 10\% uncertainty on the absolute flux scale for our ATCA observations that is typical at these wavelengths \citep{2012MNRAS.425.3137U}, and in agreement with the variations we observe of the gain calibrator flux.

\begin{figure*}
	\includegraphics[width=\textwidth]{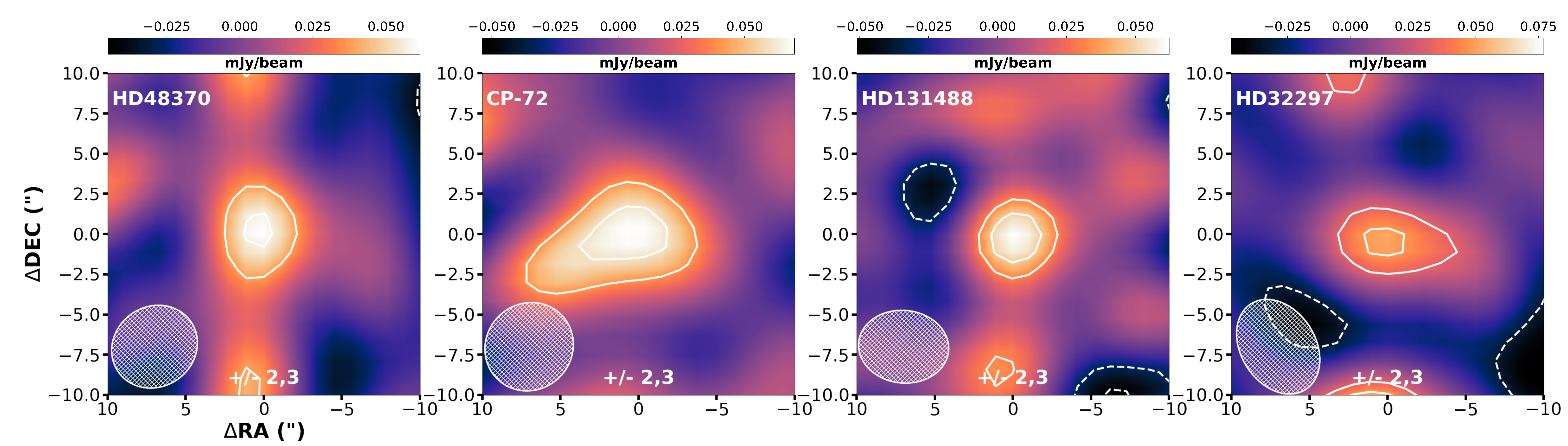}
	\caption{Images of the four debris discs detected in our 8.8-mm ATCA observations. The orientation is north up, east left. The synthesized ATCA beam FWHM extent and orientation for each observations is represented by the shaded ellipse in the bottom left of each panel. Contours are $\pm2$ and 3-$\sigma$, with negative contours denoted by dashed lines.}
	\label{fig:survey_images}
\end{figure*}

After calibrating the data, images at 34~GHz were produced using robust weighting of 2 to achieve natural weighting and retain maximum detectable flux. The dirty images were cleaned to $5\sigma$ (5 times the RMS noise level) using the \textit{clean} task and the beam was restored using the \textit{restor} task. The resulting images for our four sources are presented in Figure \ref{fig:survey_images}. HD~48370, HD~131488, and HD~32297 exhibit a north-south alignment of residual emission peaks. This is an artefact from clean due to poorly sampling in uv-space. Given the marginal detection (\(\rm \sim 3 \sigma\)) and beam size with respect to the \(\rm 3 \sigma\) contours, the discs are consistent with being unresolved and the flux density is calculated using the \textit{imfit} task (Table~\ref{tab:atca_results}) with the \textit{source} parameter set to "point". This fits a Gaussian with a width equal to the point-spread-function. We are unable to confirm the presence of stellar emission in our detections. To check for other forms of long wavelength emission in our sources, either resolved observations or temporal monitoring is required \citep{2017MNRAS.466.4083U}. However, there was a lack of strong, short (<15 min) flares that would present as obvious deviations in the amplitude vs. time relation for each target.

\section{Results}

\subsection{Spectral Energy Distributions} \label{sect:sed}
We combine our four newly acquired ATCA 8.8~mm flux measurements with  photometry from the literature to derive flux density distributions. The photometry comes from a wide variety of sources, including St\"omgren $uvby$ \citep{2015A&A...580A..23P}, Gaia \citep{2018A&A...616A...1G}, 2MASS \citep{2012yCat.2281....0C}, \textit{WISE} \citep{2010AJ....140.1868W}, \textit{Herschel} PACS  \citep[][using our own PSF photometry]{2018MNRAS.475.3046S}, and various (sub-)mm papers (see Table \ref{tab:results} for references). We then simultaneously fit stellar and disc components as described by \citet{2019MNRAS.488.3588Y}; we use \textsc{Phoenix} models \citep{2012IAUS..282..235A} for the stellar photosphere component, and a `modified' blackbody function for the disc. The modified blackbody is simply a normal Planck function $B_{\nu}(T)$ where the emission is blackbody-like and is multiplied by an additional factor \(\rm (\lambda_0 / \lambda)^{-\beta} \) beyond a fitted `break' wavelength $\rm \lambda_0$ where the slope becomes steeper. The spectral slope at long wavelengths (i.e. beyond \(\rm \lambda_0\)) is therefore $\rm 2-\beta$. Where a significantly better fit is found, two disc components are used \citep[e.g.][]{2009ApJ...701.1367C, 2014MNRAS.444.3164K}, though here we are focused on the parameters of the cooler component which contributes to the sub-mm and mm flux. The best fit models for HD~48370 and CPD-72 suggest a single dust belt is present at 39~K and 45~K respectively, whereas for HD~131488 and HD~32297  warm and cold belts were required at 414~K and 87~K, and 225~K and 82~K, respectively. Our targets have fairly shallow mm-wave slopes, meaning that the fitted values of \(\rm \beta \) are close to zero and \(\rm \lambda_0\) is essentially unconstrained. As a result, we do not quote the results for \(\rm \lambda_0\) in this manuscript. We assume a 10\% uncertainty on our derived temperatures which is generous, albeit has a minimal effect on $q$ and is contained within the flux errors associated with the spectral mm index.

\begin{figure*}
	\includegraphics[width=\textwidth]{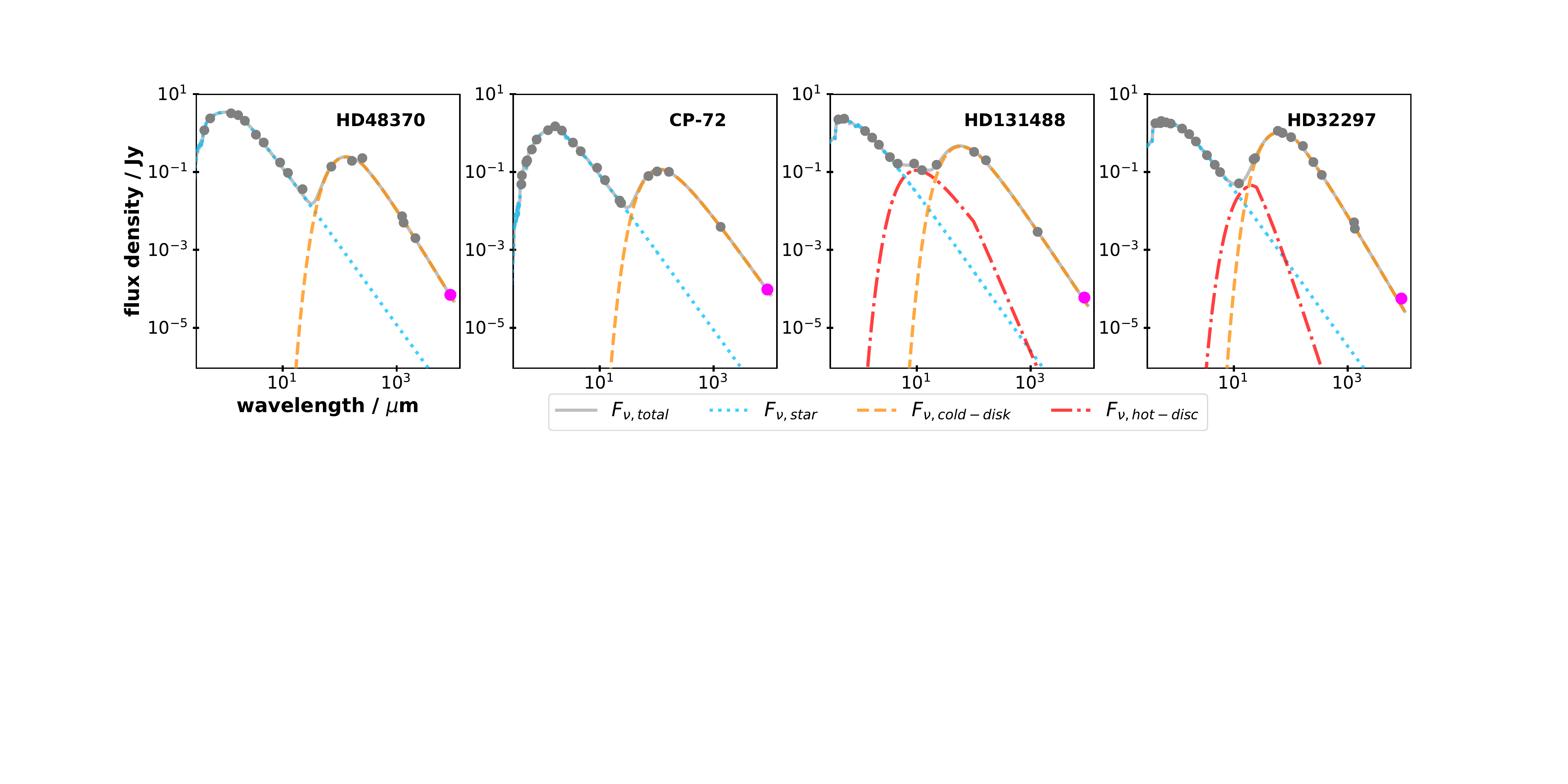}
	\caption{The spectral energy distributions of our four PLATYPUS debris discs. The magenta marker denotes the ATCA 8.8~mm flux, grey markers represent fluxes from the literature. Circle markers indicate detections above \(\rm 3\sigma\). The total flux, stellar photosphere, cold disc emission, and hot disc emission are represented by a hard grey line, dotted blue line, dashed orange line, and dash-dot red line respectively.}
	\label{fig:seds}
\end{figure*}

\subsection{Determining the Grain Size Distribution Parameter}
We derive the power law index of the dust grain distribution $q$ in an identical fashion to \citet{2016ApJ...823...79M}. Briefly, this involves calculating the slope of the Planck function, \(\alpha_{\rm Pl}\), for our 8.8~mm fluxes and the next longest wavelength, which for our sample is at 1.3~mm:
\begin{equation}
 \alpha_{\rm Pl} = \left| \frac{\log(B_{\nu_1}/B_{\nu_2})}{\log(\nu_1/\nu_2)} \right|
\end{equation}
where \(\rm B_{\nu}\) is the Planck function at the dust temperature \(T_{\rm d}\) (taken as the fitted temperature of the cool disc component in the SEDs) and \(\nu_{1,2}\) are the frequencies of our two longest wavelength observations. The mm spectral index can then be calculated via \( \alpha_{\rm mm} = |\log(F_{\nu_1}/F_{\nu_2}) / \log(\nu_1/\nu_2)|\). Assuming both flux measurements are in the Rayleigh-Jeans limit, $q$ is then derived through the relation:
\begin{equation} \label{eqn:q}
q = \frac{\rm \alpha_{mm}-\rm \alpha_{Pl}}{\beta_{\rm s}}+3 
\end{equation}
\citet{2006ApJ...636.1114D} analytically derives the $q$-$\beta$ relation (where $q$ is referred to as $p$ in the manuscript) for grains in protoplanetary discs where \(\beta_{\rm s}\) is the dust opacity spectral index of small particles. This relationship is valid for values of $q$ in the range of 3--4 which are typical for debris discs. The value of \(\beta_{\rm s}\) is dependent on the assumed material composition of the dust. \(\beta_{\rm s}\) can vary from 1.3--2 \citep{1994A&A...292..641J, 2004tcu..conf..213D}, however the variation in \( \beta_{\rm s}\) is comparable to the uncertainties on $q$ derived from our analysis. We therefore assume that our grains are composed of astronomical silicate where \(\beta_{\rm s} = 1.8\), consistent with many other debris disc studies \citep{2016ApJ...823...79M,2017MNRAS.468.2719M}. 

Table \ref{tab:results} presents $q$ values for our four sources calculated using equation \ref{eqn:q}, along with 18 other debris discs from the literature (see the references in Table \ref{tab:results}). Due to an over subtraction during the removal of stellar emission, \citet{2016ApJ...823...79M} estimates a lower limit for the disc emission in AU Mic. This results in a counter-intuitive upper limit of the $q$ value. 


\begin{table*}
 \centering
\caption{The complete debris disc sample with sub-mm to mm observations. Our four new PLATYPUS sources with ATCA 8.8~mm fluxes are in bold. $\lambda_{1_{\rm mm}}$ and $\lambda_{2_{\rm mm}}$ are the wavelengths used to calculate the millimetre spectral slope $\alpha_{\rm mm}$ (flux densities are presented in Table \ref{tab:disc_c}). $L_{\star}$ and $T_{\rm d}$ (the dust temperature of the cold component) for our four new sources is derived from the SED modelling. $\alpha_{\rm Pl}$ and $q$ are calculated using equations 1 and 2 respectively. This table has been adapted from \citet{2020A&A...641A..75L}.}
\label{tab:results}
\begin{adjustbox}{width=1\textwidth}
\begin{tabular}{@{}lcccccccccc@{}}
\toprule
Source & \begin{tabular}[c]{@{}l@{}}$L_{\star}$\\ {[}$L_{\odot}${]}\end{tabular} & Ref. & \(\rm\lambda_{1,mm}\) & \(\rm\lambda_{2,mm}\) & \(\rm\alpha_{mm}\) & Ref. & \begin{tabular}[c]{@{}l@{}}$T_{\rm d}$\\ {[}K{]}\end{tabular} & \(\rm\alpha_{Pl}\) & $q$ & Ref. \\ \midrule
AU Mic &            0.1 &   2 &     1.3 &           9.0 &   < 2.46 &                2 &    26 & 1.87 & < 3.33 & 13 \\
\textbf{CPD-72 2713} & 0.19 &   1 &     \(\rm 1.3^a\) & 8.9 & 1.95 \(\rm\pm\) 0.17 & 1 &   45 & 1.94 \(\rm\pm\) 0.12 & 3.01 \(\rm\pm\) 0.09 & 1 \\
\(\rm\epsilon\) Eri & 0.3 & 2 &     1.3 &           7.0 &   > 2.39 &                9 &    26 & 1.87 & > 3.29  & 13  \\
HD 61005 &          0.5 &   2 &     1.3 &           9.0 &   2.49 \(\rm\pm\) 0.08 &  2 &    30 & 1.89 & 3.33 \(\rm\pm\) 0.04 & 13  \\
\textbf{HD 48370} & 0.77 &  1 &     \(\rm 1.3^a\) & 8.9 &   2.26 \(\rm\pm\) 0.14 & 1 &     39 &  1.93 \(\rm\pm\) 0.14 & 3.18 \(\rm\pm\) 0.08 & 1 \\
HD 107146 &       1.0 &   2 &     1.25 &          7.0 &   2.55 \(\rm\pm\) 0.11 &  10 &    27 & 1.88 & 3.37 \(\rm\pm\) 0.06 & 13  \\
HD 377 &            1.0 &   2 &     0.87 &          9.0 &   > 2.39 &                2 &  42 & 1.92 & > 3.26 & 13  \\
HD 105 &            1.2 &   3 &     0.87 &          9.0 &   2.41 \(\rm\pm\) 0.16 &  3    & 33 & 1.90 & 3.28 \(\rm\pm\) 0.09 & 13  \\
\(\rm q^1\) Eri &   1.2 &   2 &     0.87 &          7.0 &   2.94 \(\rm\pm\) 0.10 &  2    & 33 & 1.90 & 3.58 \(\rm\pm\) 0.06 & 13  \\
HD 104860 &         1.4 &   2 &     1.3 &           9.0 &   3.08 \(\rm\pm\) 0.23 &  2    & 31 & 1.89 & 3.66 \(\rm\pm\) 0.13 & 13  \\
HD 15115 &        3.3 &   2 &     1.3 &           9.0 &   2.75 \(\rm\pm\) 0.15 &  2    & 40 & 1.92 & 3.46 \(\rm\pm\) 0.08  & 13  \\
HD 181327 &         3.3 &   2 &     1.3 &           7.0 &   2.38 \(\rm\pm\) 0.05 &  10   & 42 & 1.92 & 3.26 \(\rm\pm\) 0.03 & 13  \\
HR 8799 &           5.4 &   4 &     1.3 &           9.0 &   2.41 \(\rm\pm\) 0.17 &  11      & 40 & 1.92 & 3.27 \(\rm\pm\) 0.10  & 11 \\
\(\rm\eta\) Crv &   6.6 &   5 &     0.85 &          9.0 &   2.10 \(\rm\pm\) 0.07 &  5,12    & 38 & 1.91 & 3.11 \(\rm\pm\) 0.04 & 13  \\
\textbf{HD 32297} & 8.2 &     1 &     \(\rm 1.3^a\) & 8.9 &   2.11 \(\rm\pm\) 0.22 & 1    & 82 & 1.97 \(\rm\pm\) 0.07 & 3.07 \(\rm\pm\) 0.12 & 1 \\
HD 95086 &          8.6 &   2 &     1.3 &           7.0 &   2.41 \(\rm\pm\) 0.12 &  2    & 35 & 1.91 & 3.27 \(\rm\pm\) 0.07 & 13 \\
\(\rm\beta\) Pic &  8.7 &   2 &     0.87 &          7.0 &   2.81 \(\rm\pm\) 0.10 &  10     & 49 & 1.93 & 3.49 \(\rm\pm\) 0.06 & 13  \\
HD 131835 &         10.5 &  6 &     0.87 &          9.0 &   2.17 \(\rm\pm\) 0.13 &  12    & 56 & 1.94 & 3.13 \(\rm\pm\) 0.07 & 13 \\
\textbf{HD 131488}& 12.8 &     1 &     \(\rm 1.3^a\) &   8.9 & 2.05 \(\rm\pm\) 0.17 & 1    & 87 & 1.97 \(\rm\pm\) 0.06 & 3.04 \(\rm\pm\) 0.09 & 1 \\
Formalhaut &        16 &    2 &     1.3 &           7.0 &   2.70 \(\rm\pm\) 0.17 &  2    & 50 & 1.93 & 3.43 \(\rm\pm\) 0.09 & 13  \\
49 Ceti &           20 &    2 &     0.85 &          9.0 &   2.76 \(\rm\pm\) 0.11 &  2     & 64 & 1.95 & 3.45 \(\rm\pm\) 0.06 & 13  \\
HR 4796 A &         27 &    7,8 &   0.85 &          9.0 &   > 2.73 \(\rm\pm\) 0.10& 12    & 73 & 1.96 & 3.43 \(\rm\pm\) 0.06 & 13  \\ \bottomrule
\end{tabular}
\end{adjustbox}
    {\centering \textbf{Notes:} \(^{a}\)\(\rm\lambda_1\) fluxes are taken from \citet{2020AJ....159..288M} (CPD-72), this work (HD 48370), \citet{2018ApJ...869...75M} (HD32297), and \citet{2017ApJ...849..123M} (HD 131488). \\ \textbf{References} (1) this work, (2) \citet{2016ApJ...823...79M}, (3) \citet{2018ApJ...869...10M}, (4) \citet{2017MNRAS.470.3606H}, (5) \citet{2017MNRAS.465.2595M}, (6) \citet{2015ApJ...802..138H} (7) \citet{1999A&AS..137..273G}, (8) \citet{2016A&A...595A...1G,2018A&A...616A...1G}, (9) \citet{2015ApJ...809...47M}, (10) \citet{2015ApJ...813..138R}, (11) \citet{2018ApJ...855...56W}, (12) \citet{2017MNRAS.468.2719M}, (13) \citet{2020A&A...641A..75L}. \par}
\end{table*}

\section{Discussion}
\begin{figure*}
	\includegraphics[width=\textwidth]{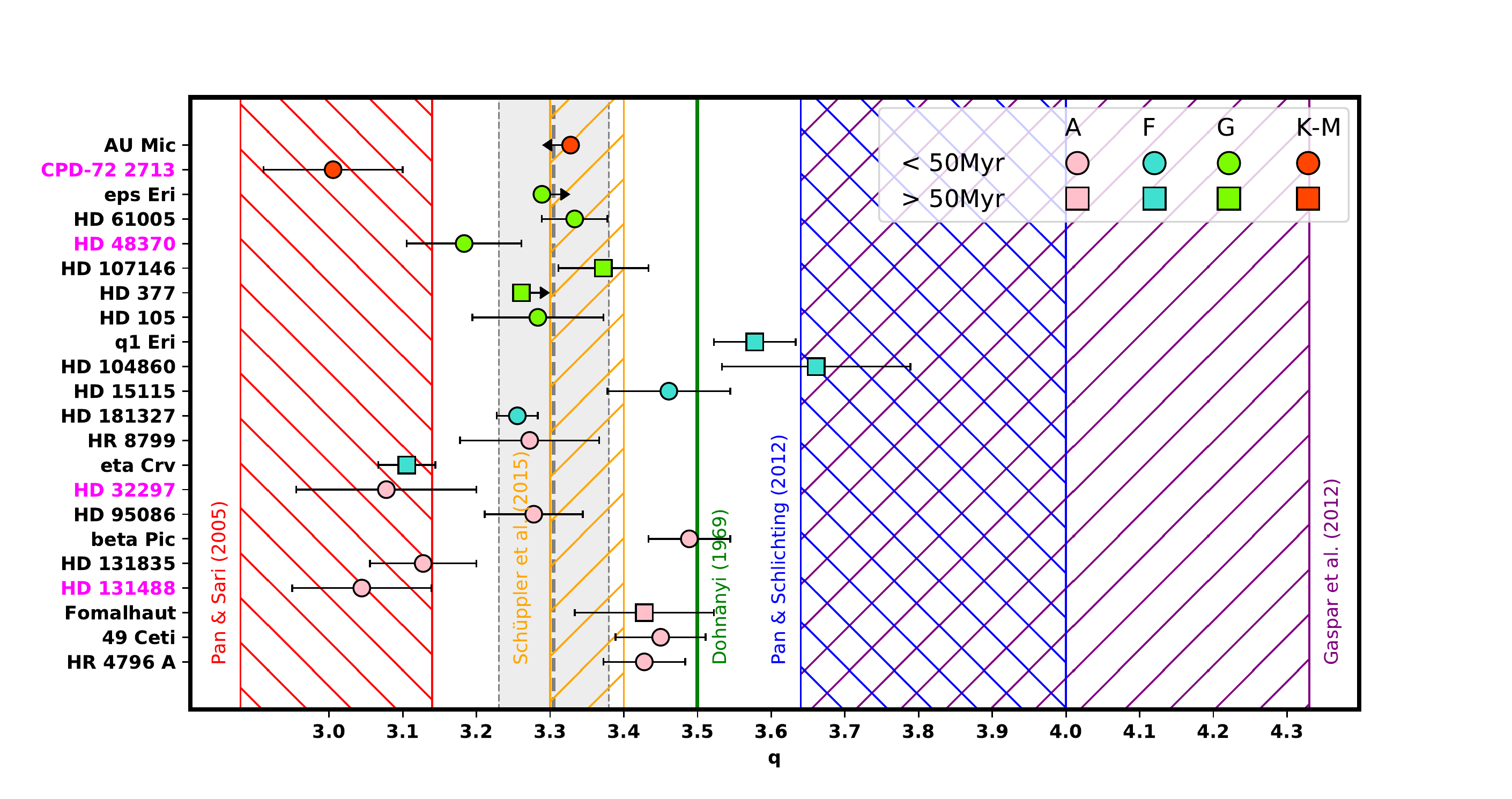}
	\caption{Distribution of grain size distribution power-law index $q$ for the 22 debris disks. The labels for our four sources with new ATCA 8.8~mm flux are presented in magenta, while other sources are labelled in black as taken from \citet{2020A&A...641A..75L} (see references within). The mean weighted $q$ value and associated uncertainty (<$q$> $= 3.31 \pm 0.07$) is shown by the grey region. Sources are ordered by their stellar luminosity from top to bottom (see Table \ref{tab:results} for specific luminosity values). The solid lines and dashed regions indicate different model predictions: ``rubble pile'' planetesimals not dominated by material strength (\citet{2005Icar..173..342P}; red), results of the ACE numerical model for AU Mic (\citet{2015A&A...581A..97S}; yellow), the classic \citet{1969JGR....74.2531D} result (green), numerical results of \citet{2012ApJ...754...74G} (purple), and incorporating a size-dependent velocity distribution (\citet{2012ApJ...747..113P}; blue). The markers are differentiated by spectral type and age (see Table \ref{tab:disc_c} for values and references). For spectral types A, F, G, and K-M the markers have colours red, green, blue, and pink respectively. Targets with ages less than/greater than 50~Myr are represented by circle/square markers respectively.}
	\label{fig:q}
\end{figure*}

Figure \ref{fig:q} shows the distribution of $q$ for 22 debris discs, the largest sub-mm to mm comparison sample to date. We include both the spectral type and age in Figure \ref{fig:q}, and find slightly lower correlations between the $q$ value and stellar properties in comparison to findings presented in \citet{2016ApJ...823...79M} (for a slightly smaller sample of 15 debris discs). For our sample of 22 debris discs there does not appear to be any correlation between age and the grain size power law. Separating our sample at 50Myr (see Table \ref{tab:disc_c} for target ages), a Kolmogorov-Smirnov (K-S) test estimates a probability of 62\% that these two populations are drawn from the same distribution. However, there is a tentative trend with spectral type (slightly weaker than that found by \citet{2016ApJ...823...79M}). Separating our sample by A-F stars and G-(K-M) stars, a K-S test estimates a probability of 23\% that these two populations are drawn from the same distribution. This suggests that stars with later spectral types may exhibit shallower grain size distributions, as previously seen in \citet{2015MNRAS.454.3207P}. However, since we increase the percentage of later type stars in our sample compared to \citet{2016ApJ...823...79M} and our correlation decreases, we suggest that as the later star debris disc sample becomes more populated it's likely that the trend between $q$ and the spectral type will disappear. We also include $q$-value model predictions from \citet{1969JGR....74.2531D}, \citet{2005Icar..173..342P}, \citet{2012ApJ...747..113P},  \citet{2012ApJ...754...74G}, and \citet{2015A&A...581A..97S} indicated by the various hatched regions in the plot. The weighted mean $q$ value of the sample, \(\langle q \rangle = 3.31 \pm 0.07\), is in close agreement with mean weighted values from previous studies analysing a subset of our sample; \(\langle q \rangle = 3.42 \pm 0.07\) \citep{2015ApJ...813..138R}, \(\langle q \rangle = 3.36 \pm 0.02\) \citep{2016ApJ...823...79M}, and \(\langle q \rangle = 3.23 \pm 0.04\) \citep{2017MNRAS.468.2719M}. These values of $q$ closely align with numerical results from \citet{2015A&A...581A..97S} rather than the larger range (3.3--4.6) presented by \citet{2014ApJ...792...65P} from far-infrared excesses. The $q$ values derived for our new ATCA observations all fall within the range predicted by \citet{2005Icar..173..342P}. This suggests that the colliding bodies in these discs have a lower tensile strength and clumping is dominated by self-gravity, resulting in a shallower size distribution. Recently, \citet{2020A&A...641A..75L} reviewed dust material approximations (\(\beta\)) and compared them with numerical results for a number of materials \citep{1983asls.book.....B,2004CoPhC.162..113W}. They suggest that the inferred grain size distribution indexes from dust material approximations are underestimated. However when comparing a single material, astronomical silicate, their numerical values for $q$ (seen in their Fig. 14) are contained within our errors in Figure \ref{fig:q}.

\begin{figure*}
	\includegraphics[width=\textwidth]{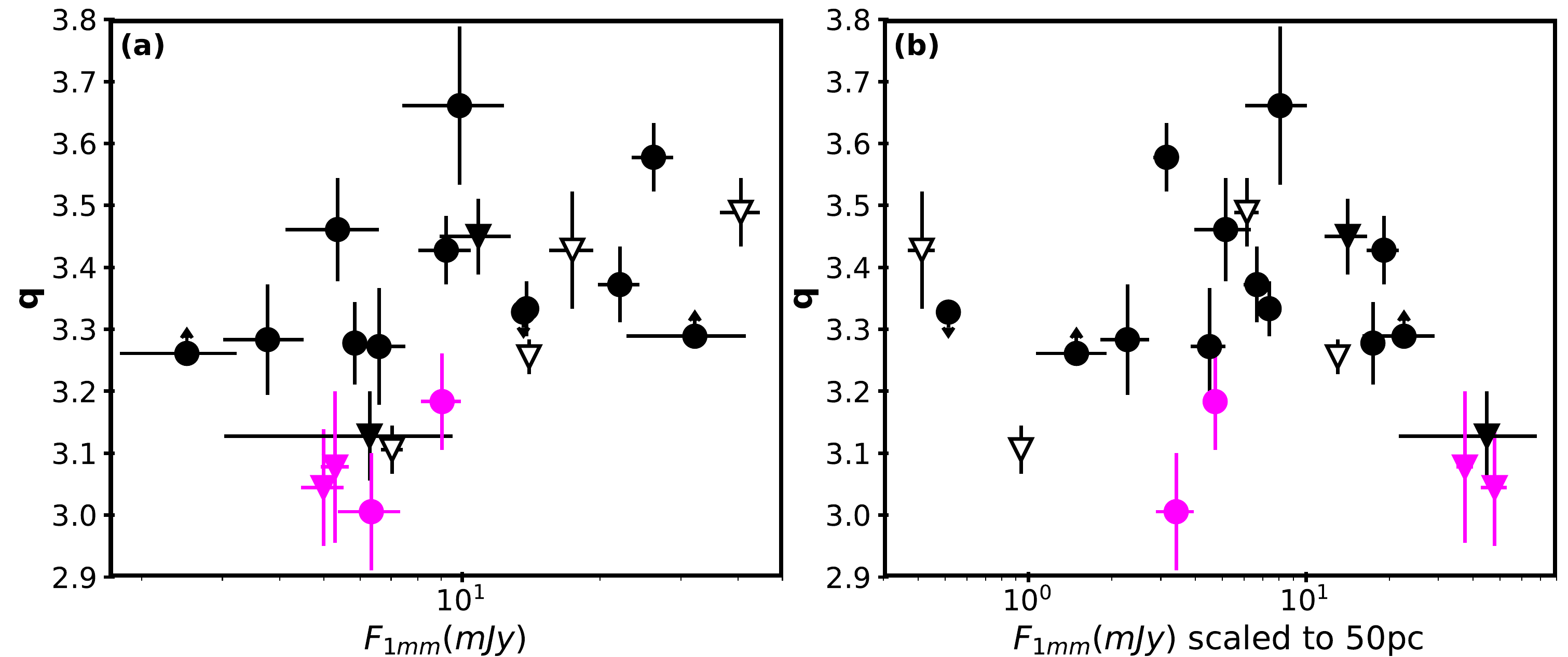}
	\caption{The grain size distribution parameter $q$ as a function of (a) the interpolated 1~mm flux, and (b) the interpolated 1~mm flux scaled to a distance of 50~pc. Distances are taken from Gaia DR2 \citep{2018A&A...616A...1G}. Our four new ATCA 8.8~mm sources are labelled by magenta circles, and discs taken from the literature are labelled by black circles. Gas-rich sources are represented by closed triangles, and those that are gas-poor are represented by open triangles. The Pearson correlation for this relation for (a) is 0.57 and for (b) is -0.22.}
	\label{fig:qvsflux}
\end{figure*}

As expected, the addition of our four new sources to the catalogue of debris discs has little impact on the  weighted mean $q$ value. However, all four targets populate the lower range of $q$ values seen to date. In Figure \ref{fig:qvsflux}, we present the grain size distribution as a function of the interpolated 1~mm flux (Fig.\ref{fig:qvsflux}a) and the interpolated 1~mm flux scaled to 50~pc (Fig.\ref{fig:qvsflux}b). Using the \textsc{scipy.stats.pearsonr} function, we find a moderate correlation between $q$ and the interpolated flux with a Pearson coefficient of 0.57. After scaling by the distance, this trend is reversed and the correlation is weaker with a Pearson coefficient of -0.22. The absence of discs in the lower-right region of Figure \ref{fig:qvsflux}a can be attributed to that fact that targets with high fluxes and shallower mm-slopes are defined as protoplanetary discs which are not included in debris disc surveys \citep[e.g. see \(\rm \alpha_{mm}\) values presented in ][]{2007A&A...462..211L, 2021MNRAS.502.5779N}. 

For optically thin debris discs, the size distribution index is related to the physics of grain collisions which may be influenced by the presence of gas. \citet{2016ApJ...828...25L} found that CO-rich systems in their sample of 12 resolved debris discs contain grain sizes on lower end of the size distribution. There are eight CO-bearing debris disks in our sample: \(\rm \beta\)~Pic \citep{2017MNRAS.464.1415M}, Fomalhaut \citep{2017ApJ...842....9M}, \(\rm \eta\)~Crv \citep{2017MNRAS.465.2595M} and HD~181327 \citep{2016MNRAS.460.2933M} contain relatively small amounts of CO gas (\(\rm M_{CO} < 10^{-4} M_\oplus \), represented by open triangles in Figure~\ref{fig:qvsflux}), whereas 49~Ceti \citep{2019ApJ...884..108M}, HD~32297 \citep{2019ApJ...884..108M}, HD~131488 \citep{2017ApJ...849..123M} and HD~131835 \citep{2015ApJ...814...42M} are CO-rich (\(\rm M_{CO} > 0.01 M_\oplus \), represented by closed triangles in Figure \ref{fig:qvsflux}). However we find no conclusive trend, although gas-rich systems have marginally lower $q$ values than gas-poor disks. This could be the result of either a gas-rich debris system preventing blow out and retaining smaller grains via gas drag, or frequent grain collisions that give rise to excess gas as well as a cascade of small dust grains \citep{2016ApJ...828...25L}. After scaling the flux (Fig. \ref{fig:qvsflux}b), it becomes apparent that there may exist two distributions of $q$ in our sample. These include a  broad distribution (for $q \sim 3.2-3.7$) of $q$ for "typical" debris discs (gas-poor/non-detection) and a lower distribution (for $q < 3.2$) for bright gas-rich discs. To check this, we use a Fisher's exact test on our data from Figure \ref{fig:qvsflux}b (a Fisher's test exactly calculates the significance of there existing two distinct distributions in a sample). We first separate the groups at $q$=3.2 and categorise the targets above and below 30~mJy (effectively separating our gas-rich discs and gas-poor/non-detection discs based on brightness), we obtain a significance of $p=0.013$. If we instead categorise for gas-rich and gas-poor (which doesn't take the brightness into account) we obtain a lower significance of $p=0.04$. Thus, there is some evidence that bright (in absolute terms) gas-rich debris discs tend to contain lower $q$ values (the first test with $p=0.013$). Or alternately given that our systems are faint in apparent flux, it is possible that an observational bias exists, especially since the Pearson coefficient becomes weaker once we scale the flux with the distance. For a fixed FIR/mm flux \citep[e.g. the DUNES survey, ][]{2013A&A...555A..11E} the discs with the steepest size distribution will result in \(\rm \sim\)cm fluxes that are below the detection limit. It is quite plausible that there exist many discs lurking below current sensitivity levels \citep{2015ApJ...801..143M} and, as a result, the current mean weighted $q$ value would be biased towards lower values. If there indeed does exist a population of lower $q$ discs then the mean weighted grain size power law index would shift closer towards the analytical models presented by \citet{2005Icar..173..342P} ($2.88<q<3.14$) where colliding bodies are held together by self-gravity and contain relatively low tensile strengths.

\section{Conclusion}
In this work we present new ATCA 8.8~mm observations of four debris discs and combine them to present the largest sample to date of 22 debris discs for which the grain size distribution power-law index $q$ can be calculated to provide insights into the planetesimal populations in these discs. Our key findings are:

\begin{enumerate}
\item We present the longest wavelength observations to date of HD~48370, CPD-72~2713, HD~131488, and HD~32297 at 8.8~mm, and find that the $q$ value of these sources are all quite low ($q < 3.2$), suggesting that the colliding bodies in these discs have a lower tensile strength and clumping is dominated by self-gravity.

\item For the entire sample of 22 debris discs, we evaluate a weighted mean value of the sample, <$q$> = 3.31, consistent with analytical and numerical predictions for collisional cascade models.

\item With a larger sample (22 compared to 15 discs) we find that the tentative trend between $q$ and the spectral type becomes weaker in comparison to findings from \citet{2016ApJ...823...79M}.

\item We suggest possibility of two distributions of $q$; a broad distribution (where $q \sim 3.2-3.7$) for "typical" debris discs (gas-poor/non-detection), and a lower distribution (where $q < 3.2$) for bright gas-rich discs.

\item Or alternatively, we suggest an observational bias may be present between the grain size distribution parameter and absolute flux which is likely attributed to the detection rates of faint debris discs at \(\rm \sim\)cm wavelengths.

\end{enumerate}

\section*{Acknowledgements}
The authors thank the referee for their constructive comments and suggestions. We thank Elodie Thilliez for useful discussions regarding our ATCA observing proposal. B.J.N. is supported by an Australian Government Research Training Program (RTP) Scholarship. G.M.K. is supported by the Royal Society as a Royal Society University Research Fellow. J.P.M. acknowledges research support by the Ministry of Science and Technology of Taiwan under grants MOST107-2119-M-001-031-MY3, MOST107-2119-M-001-031-MY3, and MOST109-2112-M-001-036-MY3, and Academia Sinica under grant AS-IA-106-M03. A.M. and P. \'A. acknowledge support from the grant KH-130526 of the Hungarian NKFIH. The Australia Telescope Compact Array is part of the Australia Telescope which is funded by the Commonwealth of Australia for operation as a National Facility managed by CSIRO. This research has made use of NASA's Astrophysics Data System. The National Radio Astronomy Observatory is a facility of the National Science Foundation operated under agreement by the Associated Universities, Inc. ALMA is a partnership of ESO (representing its member states), NSF (USA) and NINS (Japan), together with NRC (Canada) and NSC and ASIAA (Taiwan) and KASI (Republic of Korea), in cooperation with the Republic of Chile. The Joint ALMA Observatory is operated by ESO, AUI/ NRAO and NAOJ. This paper makes use of the following ALMA data: ADS/JAO.ALMA\#2016.2.00200S

\textit{Software}: The {\sc Miriad} package \citep{1995ASPC...77..433S}, Python version 3.7, {\sc Astropy} \citep{2013A&A...558A..33A,2018AJ....156..123A}, {\sc SciPy} \citep{2020SciPy-NMeth}, {\sc NumPy} \citep{2020NumPy-Array}, and {\sc matplotlib} \citep{Hunter:2007}.

\textit{Facilities:} Atacama Large Millimeter/sub-millimeter Array, Australian Telescope Compact Array

\section*{Data Availability}
The ATCA observational data used in this paper is available from the Australian National Telescope Facility Archive at \url{https://atoa.atnf.csiro.au/} under project code C2696. The ALMA observational data is available from the ALMA science archive at \url{https://almascience.nrao.edu/aq/} under the project IDs listed in the Acknowledgements. Reduced observation data and models will be shared on reasonable request to the corresponding author.




\bibliographystyle{mnras}
\bibliography{mnras} 



\appendix
\section{Debris Disc Sample Characteristics}
\begin{table*}
 \centering
\caption{Debris Disc Sample Characteristics. Targets are ordered identically to Table \ref{tab:results} and our four new PLATYPUS sources with ATCA 8.8~mm fluxes are in bold. $r_{\rm d}$ is the outer radial extent of the dust disc. \(\rm \lambda_{1,mm}\) and \(\rm \lambda_{2,mm}\) are the flux density values referenced in Table \ref{tab:results}.}
\label{tab:disc_c}
\begin{adjustbox}{width=0.98\textwidth}
\begin{tabular}{@{}lccccccccccc@{}}
\toprule
Source & SpT & Ref. & \begin{tabular}[c]{@{}l@{}} Age \\ {[}Myr{]}\end{tabular} & Ref. & \begin{tabular}[c]{@{}l@{}}$r_{\rm d}$\\ {[}au{]}\end{tabular} & Ref. & \begin{tabular}[c]{@{}l@{}} $\rm Dist.^{a}$\\ {[}pc{]}\end{tabular}  & \begin{tabular}[c]{@{}l@{}} \(\rm \lambda_{1,mm}\) \\ {[}\(\rm mJy\){]}\end{tabular}  & Ref. & \begin{tabular}[c]{@{}l@{}} \(\rm \lambda_{2,mm}\) \\ {[}\(\rm \mu Jy\){]}\end{tabular}  & Ref. \\ \midrule
AU Mic &                M1   &  1   &   24        &   1  & 40   & 15 & 9.72  & \(\rm 7.14 \pm 0.15 \) & 34 & \(\rm >60.8 \pm 5.2 \) & 1 \\
\textbf{CPD-72 2713} &  K-M  &  2   &   24        &   11 & 140  & 16 & 36.6 & \(\rm 3.80 \pm 0.59 \) & 16 & \(\rm 95.9 \pm 16.1 \)& 35\\
\(\rm\epsilon\) Eri &   K2   &  1   &   400-800   &   1  & 69   & 17 & 41.8 & \(\rm 17.2 \pm 5.0 \) & 36 & \(\rm 66.1^{+6.9}_{-10.5}\) & 36\\
HD 61005 &              G8   &  1   &   40        &   1  & 67   & 18 & 36.4 & \(\rm 7.2 \pm 0.3 \) & 37 & \(\rm 57.3 \pm 8.6 \) & 1\\
\textbf{HD 48370} &     G8   &  3   &   20-50     &   3  & 90   & 19 & 36.1 & \(\rm 5.0 \pm 0.5 \) & 35 & \(\rm 70 \pm 10.8 \) & 35\\
HD 107146 &             G2   &  1   &   80-200    &   1  & 116  & 20 & 27.4 & \(\rm 12.5 \pm 1.3 \) & 38 & \(\rm 166.0 \pm 25.2 \)& 39\\
HD 377 &                G2   &  1   &   150       &   1  & 101   & 21 & 38.5 & \(\rm 3.5 \pm 1.0 \) & 23 & \(\rm <13.1 \pm 4.4 \) & 1\\
HD 105 &                G0   &  4   &   28        &   4  & 85   & 4  & 38.8 & \(\rm 2.0 \pm 0.4 \)  & 4 & \(\rm 42 \pm 14 \) & 40\\
\(\rm q^1\) Eri &       F9   &  1   &   4800      &   1  & 85   & 22 & 17.3 & \(\rm 39.4 \pm 4.1 \)  & 41 & \(\rm 92.6\pm 16.6 \)  & 39\\
HD 104860 &             F8   &  1   &   140       &   1  & 110  & 23 & 45.2 & \(\rm 4.4 \pm 1.1 \) & 23 & \(\rm 14.0 \pm 3.5\)  & 1\\
HD 15115 &              F2   &  1   &   21        &   1  & 97   & 24 & 49.0 & \(\rm 2.6 \pm 0.6 \) & 42 & \(\rm 12.8 \pm 4.1 \) & 1\\
HD 181327 &             F6   &  1   &   24        &   1  & 86   & 25 & 48.2 & \(\rm 7.5 \pm 0.1 \) & 25 & \(\rm 145.0\pm 19.2 \) & 39 \\
HR 8799 &               A5   &  5   &   30        &   5  & 232  & 26 & 41.2 & \(\rm 3.5\pm 0.5 \) & 26& \(\rm 32.6\pm 9.9 \) & 26\\
\(\rm\eta\) Crv &       F2   &  6   &   1000-2000 &   6  & 152  & 6  & 18.2 & \(\rm 9.2\pm 0.5 \) & 6 & \(\rm <36 \) & 40\\
\textbf{HD 32297} &     A5/6 &  7   &   15-45     &   12 & 100  & 18 & 133 & \(\rm 3.04\pm 0.21 \) & 12 & \(\rm 56.2\pm 16.7 \) & 35\\
HD 95086 &              A8   &  1   &   17        &   1  & 208  & 27 & 86.4 & \(\rm 3.1\pm 0.18 \) & 27 & \(\rm 61.9\pm 15.9 \)& 39\\
\(\rm\beta\) Pic &      A6   &  1   &   24        &   1  & 106  & 28 & 19.4 & \(\rm 60\pm 6\) & 43 & \(\rm 240.0\pm 33.2\) & 39\\
HD 131835 &             A2   &  8   &   15        &   8  & 85   & 29 & 133 & \(\rm 8.5\pm 4.4\) & 44 & \(\rm 53\pm 17\) & 40\\
\textbf{HD 131488}&     A2   &  9   &   15        &   14 & 91   & 30 & 155 & \(\rm 2.91\pm 0.31\) & 45 & \(\rm 59.5\pm 12.4\) & 35\\
Formalhaut &            A4   &  1   &   440       &   1  & 136  & 31 & 7.70 & \(\rm 27\pm 3\) & 46 & \(\rm 400\pm 64\) & 47\\
49 Ceti &               A1   &  1   &   40        &   1  & 95   & 32 & 57.1 & \(\rm 17\pm 3\) & 32 & \(\rm 25.1\pm 5.5\) & 1 \\
HR 4796 A &             A0   &  10  &   9         &   5  & 78   & 33 & 71.9 & \(\rm 14.4\pm 1.9\) & 5 & \(\rm < 63 \) & 40\\ \bottomrule
\end{tabular}
\end{adjustbox}
    {\centering \textbf{Notes:} \(^{a}\) Distances are taken from \citet{2018A&A...616A...1G}. \\ \textbf{References} (1) \citet{2016ApJ...823...79M}, (2) \citet{2006A&A...460..695T,2013ApJS..208....9P}, (3) \citet{2008hsf2.book..757T}, (4) \citet{2018ApJ...869...10M}, (5) \citet{2017MNRAS.470.3606H}, (6) \citet{2017MNRAS.465.2595M}, (7) \citet{2009ApJ...702..318D}, (8) \citet{2015ApJ...802..138H}, (9) \citet{2013ApJ...778...12M}, (10) \citet{1999A&AS..137..273G}, (11) \citet{2006A&A...460..695T,2015MNRAS.454..593B,2018MNRAS.475.2955L,2018ApJ...856...23G}, (12) \citet{2018ApJ...869...75M}, (13) \citet{2002AJ....124.1670M,2012ApJ...746..154P}, (15) \citet{2015ApJ...811..100M}, (16) \citet{2020AJ....159..288M}, (17) \citet{2017MNRAS.469.3200B}, (18) \citet{2018ApJ...869...75M}, (19) \citet{2016ApJ...826..123M}, (20) \citet{2018MNRAS.479.5423M}, (21) \citet{2016ApJ...817L...2C}, (22) \citet{2010A&A...518L.132L}, (23) \citet{2016ApJ...816...27S}, (24) \citet{2019ApJ...877L..32M}, (25) \citet{2016MNRAS.460.2933M}, (26) \citet{2018ApJ...855...56W}, (27) \citet{2017AJ....154..225S}, (28) \citet{2019AJ....157..135M}, (29) \citet{2015ApJ...815L..14H}, (30) \citet{2019MNRAS.489.3670K}, (31) \citet{2017ApJ...842....8M}, (32) \citet{2017ApJ...839...86H}, (33) \citet{2018MNRAS.475.4924K}. (34) \citet{2013ApJ...762L..21M} (35) this work, (36) \citet{2015ApJ...809...47M}, (37) \citet{2013ApJ...774...80R}, (38) \citet{2015ApJ...798..124R}, (39) \citet{2015ApJ...813..138R}, (40) \citet{2017MNRAS.468.2719M}, (41) \citet{2008A&A...480L..47L}, (42) \citet{2015ApJ...801...59M},  (43) \citet{2014Sci...343.1490D}, (44) \citet{2015ApJ...814...42M}, (45) \citet{2017ApJ...849..123M}, (46) \citet{2003ApJ...582.1141H}, (47) \citet{2012A&A...539L...6R}\par}
\end{table*}


\bsp	
\label{lastpage}
\end{document}